# Space-temporal structure of the thunderstorm ground enhancements (TGEs)

A.Chilingarian, D.Pokhsraryan, F.Zagumenov, and M.Zazyan

*A I Alikhanyan National Laboratory, Yerevan 0036, Armenia*

## Abstract

We analyzed the structure of the Thunderstorm Ground Enhancement (TGE) using a particle detector network on Aragats. We performed a statistical analysis of the particle flux enhancement time series on a nanosecond time scale using the largest TGE event on record, which occurred on May 23, 2023. Our findings confirm that the TGE combines multiple Extensive Cloud Showers (ECSs), or particle microbursts, that arrive independently and provide stable particle flux on a second-time scale. The electron accelerator, operated by the dipole that emerges in the lower part of the thundercloud, sends copious electrons and gamma rays toward the Earth's surface that sustains for minutes. The experimental results are supported by simulations of electron multiplication and acceleration in strong atmospheric electric fields.

## 1. Introduction

Observation of the intense particle fluxes on the earth's surface, so-called TGEs [1,2], and numerous detections of a few gamma rays reaching the orbiting gamma-ray observatories [3], for first glance, seem to be different physical phenomena. However, there are no fundamental differences in the origination of TGF and TGE processes. Relativistic runaway electron avalanche (RREA, [4-7]) develops in the lower dipole to generate a TGE observed by particle detector on the ground and - in the upper dipole to generate TGF 500 km higher. TGFs are the most energetic bremsstrahlung gamma rays, occasionally reaching the orbiting gamma detectors from strong equatorial thunderstorms. Fast-moving detectors 300-500 km from the source can hardly detect more than a few RREAs within several tens of microseconds. The Atmosphere-Space Interactions Monitor (ASIM) on board the International Space Station is the only one specially constructed for TGF detection and has already measured prolonged gamma-ray bursts from RREAs one following another within a time window up to 10 ms [8].

Simulations of the electron flux traversal in the strong atmospheric field revealed the copious multiplication and acceleration of electrons, forming RREAs, which reached the earth's surface and ended up as a TGE. The process's initial name, "runaway breakdown" (RB), is now changed to RREA without any relation to the breakdown initiated by the lightning flash. The basics of the runaway process are the acceleration of free electrons originating from small and large extensive air showers (EASs). In the atmosphere, there is a rather stable concentration of free electrons; those energy spectra are well known up to tens of km above the ground. Acceleration and multiplication of free atmospheric electrons become possible if the intracloud electric field is larger than the critical value, specific for the particular air density [9-11]. A vast amount of fully described TGE events are available from the Mendeley datasets [12-13] and the database of the Cosmic Ray Division of Yerevan Physics Institute.

Developing in the thunderous atmosphere, RREAs form a prolonged (sometimes up to a few ten minutes) flux of electrons and gamma rays. Thus, besides the 10-100 kA current of eV energy electrons (lightning stroke), there exists a minutes-long current of the MeV electrons in the atmosphere. A lightning flash usually terminates this current [14]. If we inverted the previous statement, we could adopt that RREA is a precursor of a lightning leader, and without RREA, the lightning flash will hardly strike the ground [15]. The lightning current flows smoothly and steadily for up to 100 ms; the RREA current comes in discreet portions. We call these avalanches extensive cloud showers, ECSs [16]; Alex Gurevich called them "microbursts" [17]. The first attempt to estimate the space-temporal structure of RREA was done on Aragats by the MAKET surface array triggers during large TGEs [2].

Analyzing the TGEs registered on Aragats on 19 September 2009 and 4 October 2010 [2,16], we demonstrated that TGE particles are more or less uniformly distributed within the TGE duration (a few minutes). To get further insight into the RREA space-time distribution, we performed a new experiment now analyzing TGE particles on the nanosecond time scale. Using a fast-synchronized data acquisition system (FSDAQ, [18]), we revealed the TGE space-time structure by detecting avalanche particles' arrival with scintillators directly attached to high-speed oscilloscopes.

## 2. Method

The TGE detection uses 18 plastic scintillators with thicknesses of 1 and 3 cm and an area of 1 square meter each commissioned by the High-Energy Physics Institute in Protvino. RF [19]. Of the 18 scintillators, 12 are part of the STAND1 network, which is located across the Aragats station. The remaining four scintillators belong to the CUBE detector and are placed inside the SKL experimental hall. All three identical units of the STAND1 network are located near three main experimental halls – MAKET, SKL, and GAMMA, see Fig.1a. Three 1-cm thick scintillators are stacked vertically, and one 3-cm thick plastic scintillator stands apart; see Figs 1b and 1c. STAND1 detector has been in operation for ten years, registering spatial distribution of more than 300 TGE at millisecond time scales. The light from the scintillator through optical spectrum-shifter fibers is passed to the photomultiplier FEU-115M. The maximum luminescence is emitted at the 420-nm wavelength, with a luminescence time of about 2.3 ns. The STAND1 detector is tuned by changing the high voltage applied to the PMT and setting the thresholds for the discriminator shaper. The discrimination level is chosen to guarantee high signal detection efficiency and maximal suppression of photomultiplier noise. The efficiency of scintillators reaches 90% and more for electron energies above 10 MeV and 2% for gamma rays with energies above 2 MeV (for the upper scintillators). The energy threshold of the upper scintillators is 0.7-0.8 MeV, and dead time is ≈0.7 μs. A 50 μs time series of STAND1 detectors, synchronized with NSEF measurements and meteorological parameters, are transferred to the Cosmic ray Division (CRD) database and are available online in graphical and numerical format via the ADEI data analyzing platform [20].

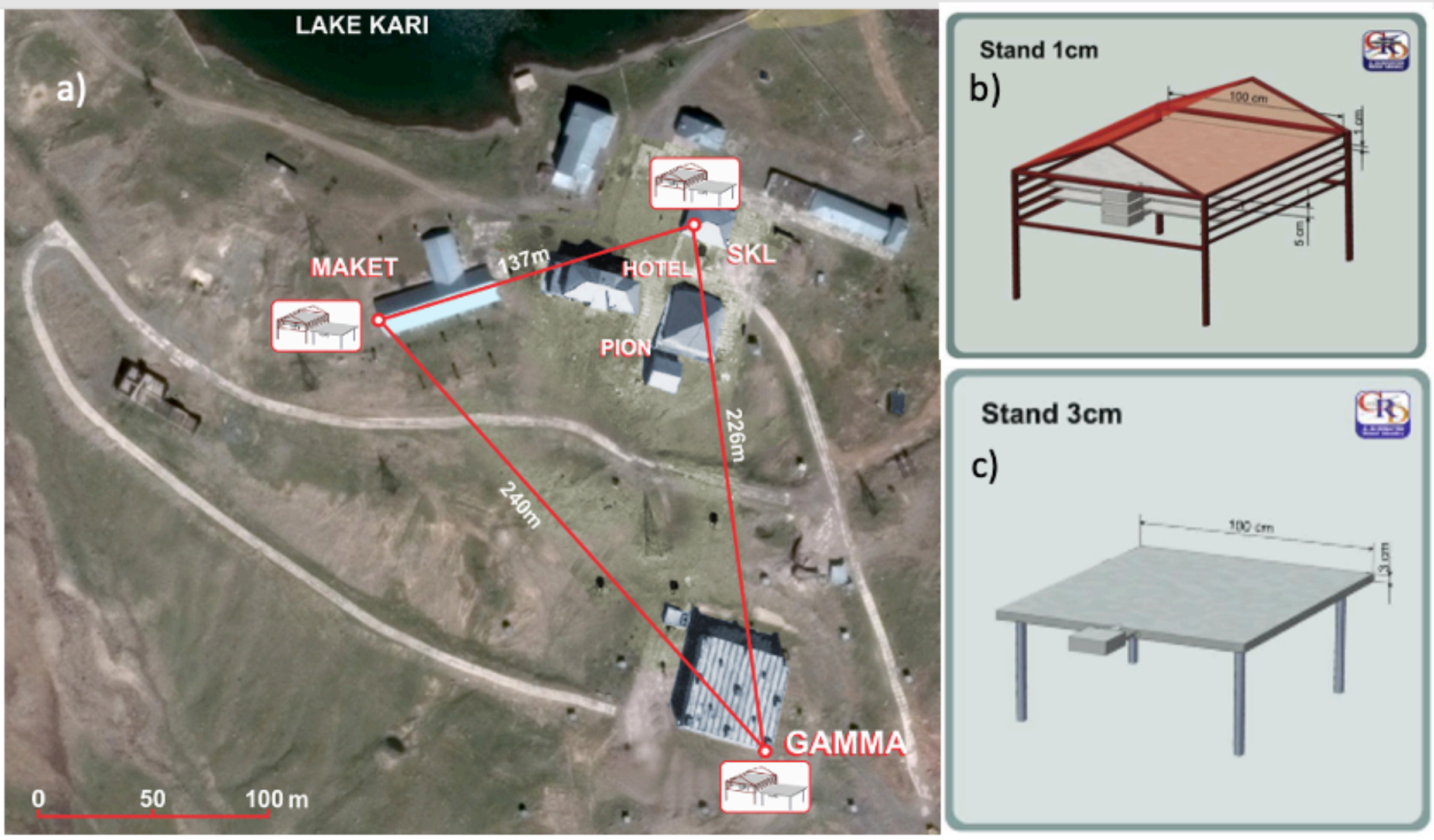

**Figure 1. a) The map of Aragats station with STAND1 network; b) Stand1 unit: vertically stacked 1 cm thick and 1 m$^2$ area plastic scintillators; c) Stand1 unit: stand-alone 3 cm thick plastic scintillator with the same area.**

Four 1 cm thick and one 3 cm thick scintillators are attached to high-speed digital oscilloscopes. A four-channel (Picoscope 6403D) and two-channel (Picoscope 5244B) digital oscilloscopes were used to measure the amplitudes of the signals from scintillators (see Fig. 2 for 4-channel Picoscope 6403D, located in the SKL experimental hall). In the MAKET experimental hall, the 3-cm thick scintillator was attached to the 2-channel oscilloscope. The record length of both oscilloscopes was 200 ms, and the sampling rate of signals from different detectors was 250 MS/s and 156.25 MS/s, corresponding to the sampling intervals of 4 ns and 6.4 ns. The typical duration (full width on half maximum, FWHM) of individual pulses from the scintillators was 25-30 ns. Thus, usually, the signal occupied several sampling intervals. One scintillator, attached to the National Instruments (NI) MyRIO board, was used to calculate the 1-second count rate. If the count rate of the 1-second time series exceeded the prechosen limit (usually set to 50% larger than the average count rate), the MyRio board produced the triggering pulse, see Fig. 2. The oscilloscope's trigger-out (synchro) pulse was relayed to the board, which produced the GPS time stamp of the record. This feature allows accurate time synchronization; the estimated resolution to meet is 100 ns. A detailed description of our fast data acquisition system based on the NI MyRIO board can be found in [21].

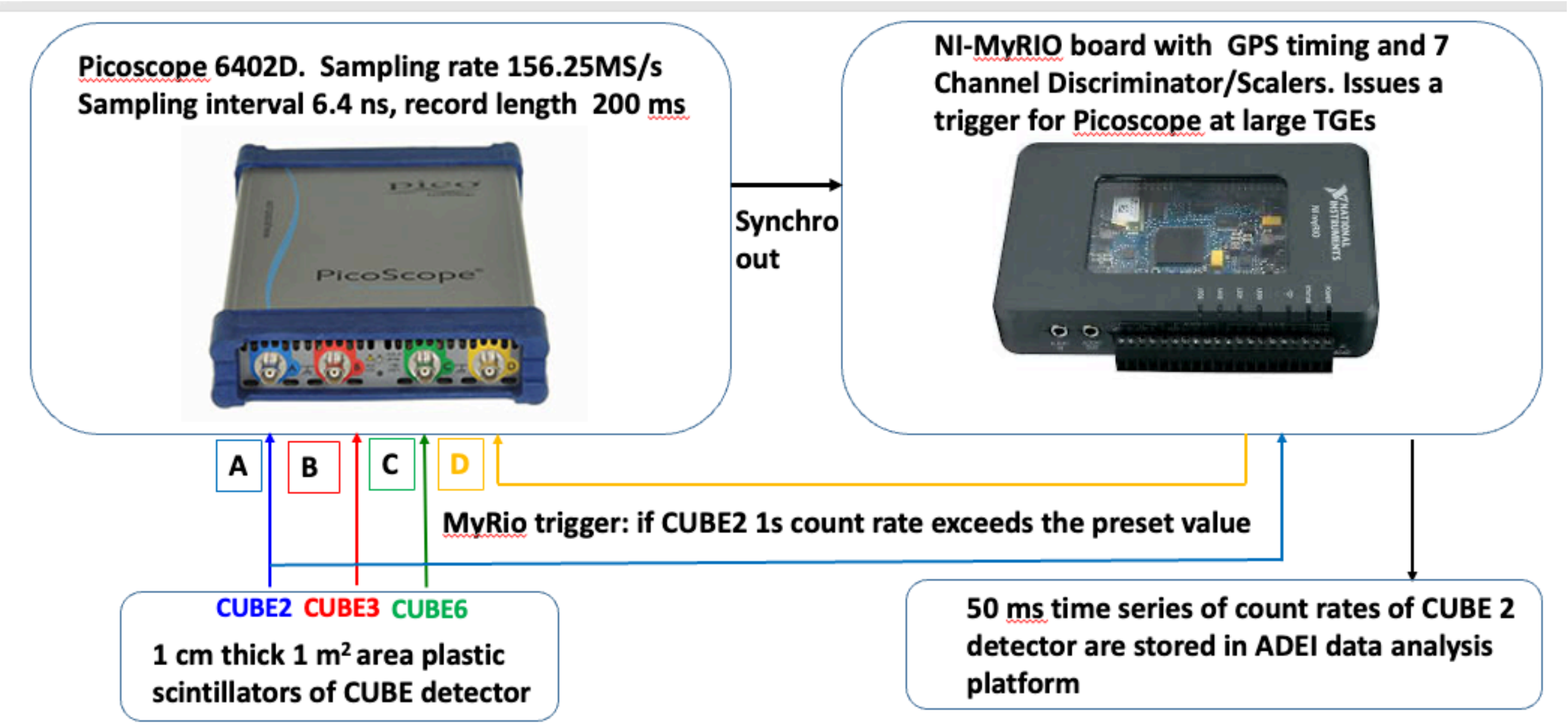

**Figure 2. Block diagram of the fast-synchronized data acquisition (FSDAQ) for the research of the space-time structure of TGEs located in the SKL experimental hall, a similar system operates in MAKET hall (two-channel Picoscope 5244B, with attached 3-cm thick plastic scintillators).**

Figure 3 displays a photo of the CUBE detector rearranged and deployed in the SKL hall. The CUBE detector was initially comprised of stacked 20 cm thick and 0.25 $m^2$ area plastic scintillators surrounded by all sides with 6 1 cm thick plastic scintillators. The main goal of the detector was gamma ray spectroscopy and estimation of the electron-to-gamma ray ratio during TGEs. In 2019, we dismounted the side and bottom detectors and put the upper (veto) scintillator directly above the 20 cm thick scintillators. Thus, the main functions of CUBE scintillators remain, and new experiments with CUBE scintillators and digital oscilloscopes started. We maintain these measurements using the CUBE 1 as a veto scintillator to reject the charged flux (the neutral flux is registered by 20 cm thick spectrometric scintillators below it).

Additionally, we consistently keep track of the signal amplitudes from CUBE detectors 2, 3, and 6. These detectors are triggered when the biggest TGEs are detected at Aragats station. CUBE 1 detector's one-minute count rate was also used to calibrate the flux measured by the 2, 3, and 6 scintillators.

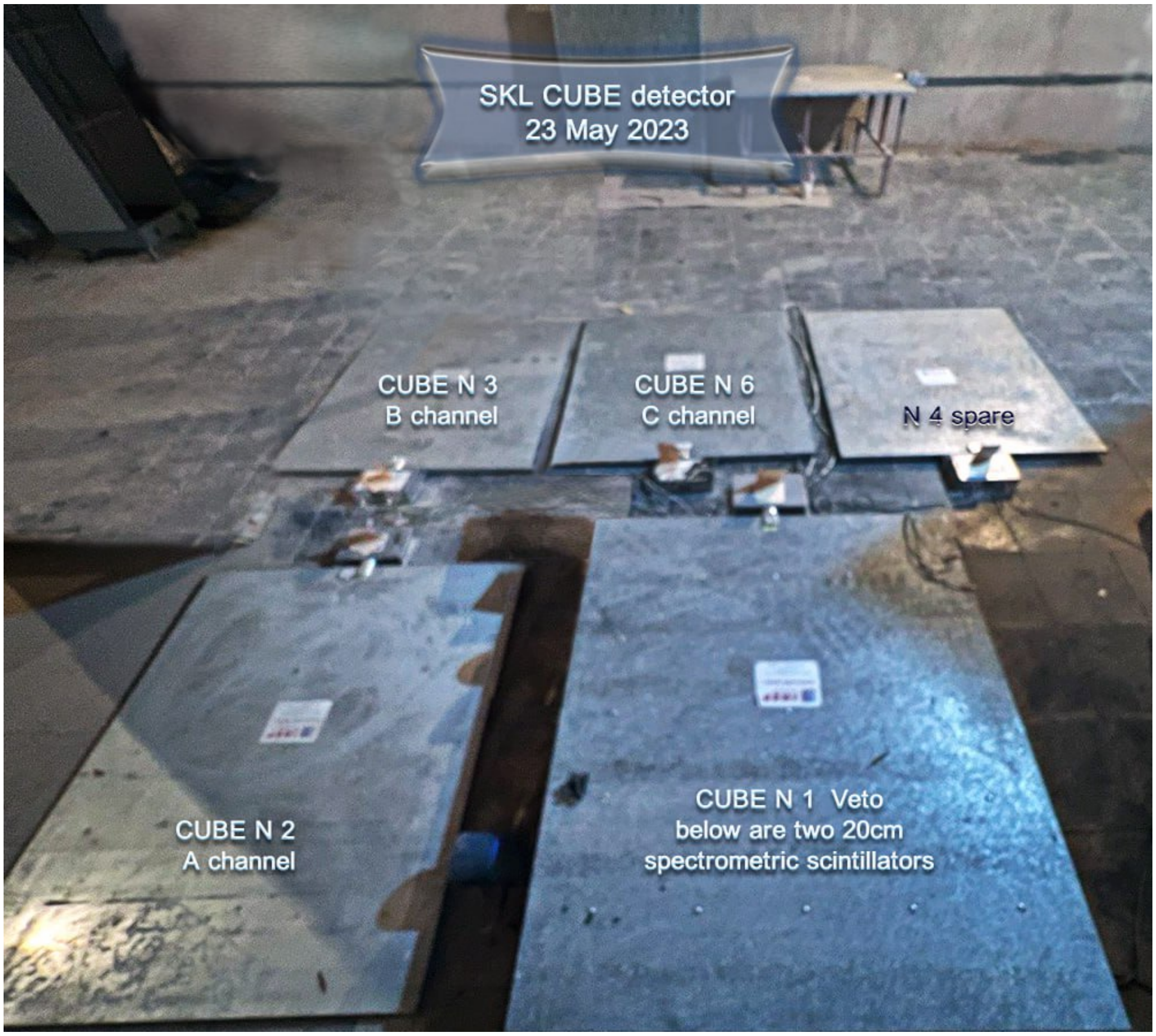

**Figure 3. The CUBE detector's scintillators are displayed in the SKL experimental hall under a 7mm tilt and a 1cm wooden roof.**

## 3. TGE occurred on 23 May 2013

TGEs comprise prominent peaks in the time series of particle fluxes originating from the electron-gamma avalanches in thunderclouds. They are observed mostly in the spring and autumn seasons. An electron accelerator in thunderclouds sends copious electrons, gamma rays, and rarely neutrons in the direction of the earth. During the last decade, particle detectors on Mt. Aragats in Armenia, Mt. Lomnicky Stit in Slovakia, Mt. Musala in Bulgaria, and now also at Mt. Zugspitze in Germany registered nearly a thousand TGEs [22,23]. Particle flux enhancement on Aragats usually counts 10-20%, never over 200%. Most TGEs and all large (> 50%) occurred in Spring and Autumn (≈80%) when the outside temperature is in the -3C° - +3C° range and clouds are very low above the Aragats research station (yellow and green colors on the histogram of Fig. 4). The standard cloud electrification due to warm air updraft and following hydrometeor tension led to the charge separation and emergence of lower dipole accelerated electrons downward. When the intracloud electric field exceeds the runaway threshold strength, an RREA process initiates a particle avalanche. We accept the tripole model of the electrostatic field of a thunderstorm according to Kuettner measurements at the Zugspitze in 1945-1949 [24]. Approximately 12% of TGE events occurred in Summer (red color) and ≈ 8% in Winter (black color). The NSEF during thunderstorms on Aragats varies from -30 – +20 kV/m, and the intracloud electric field sometimes exceeds 2.1 kV/cm (not a direct measurement, estimated from the recovered energy spectra), prolonging almost to the earth's surface.

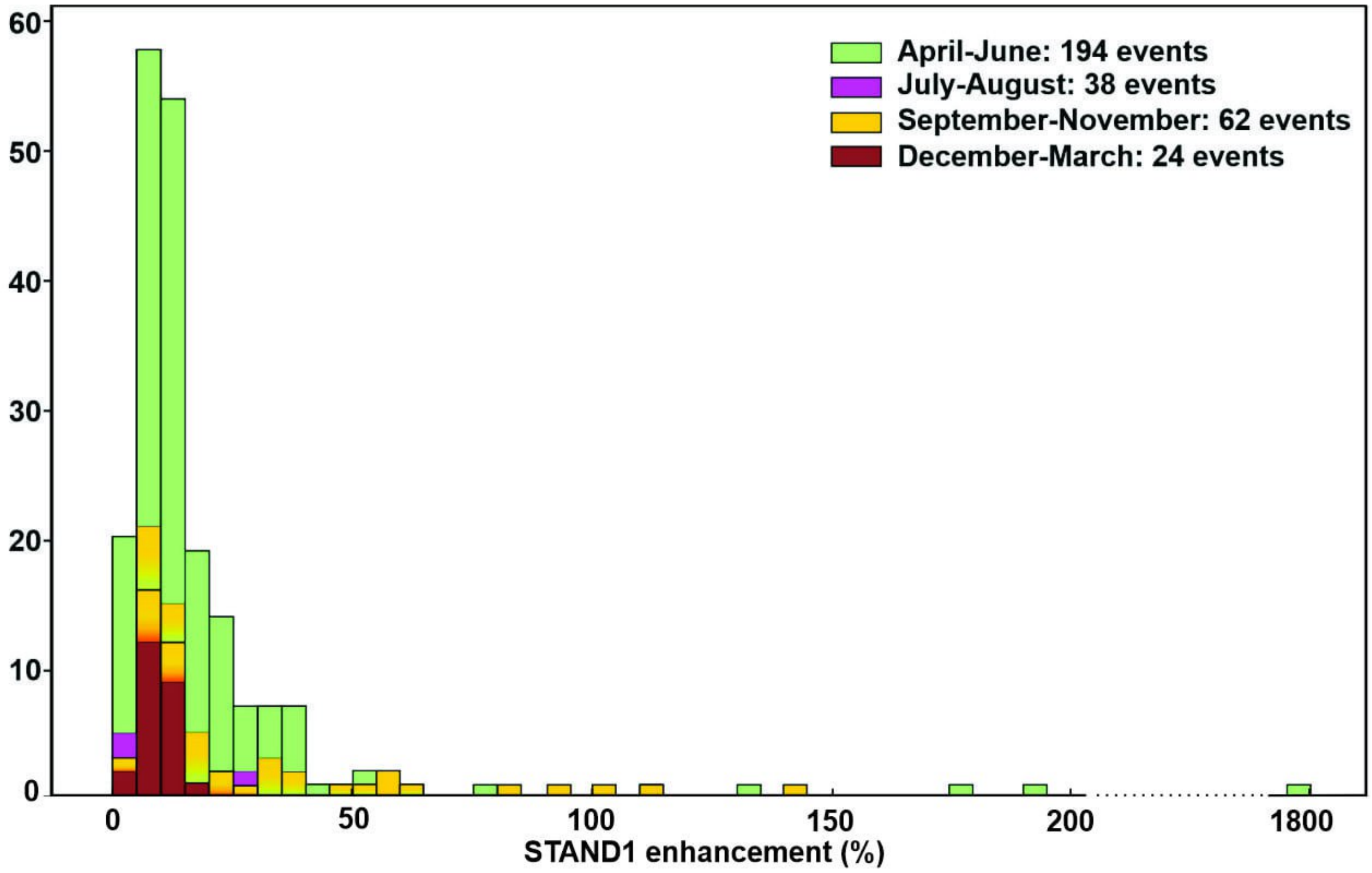

**Figure 4. The season-dependent histogram of TGE enhancements in percent. The 1 cm thick and 1 m$^2$ area plastic scintillator of the STAND1 detector (signal only in the upper scintillator of 3 stacked) was used for the relative enhancement calculation (shown in the inset).**

In Fig. 4, the histogram of the TGE occurrences is ranged according to the percent of flux enhancement. For comparative purposes, the four seasons are presented in different colors. The histogram showed 318 TGE events from 11 years (2013-2023) when the electric field sensors and weather stations were installed on Aragats. In 2008-2012, 277 TGE events were observed; however, meteorological and electricity sensors were not installed. Thus, the total number of TGE events surpasses 600. The TGEs were selected if three independent particle detectors demonstrate simultaneous peaks in the count rate time series larger than three sigma and the NSEF absolute value exceeds five kV/m.

On the right side of Figure 4, we show the enhancement counting ≈ 1500% of an extraordinarily large TGE that occurred on 23 May 2023. While such outliers are expected in an infinite number of measurements, it is unlikely to obtain such a huge TGE in only 318 trials. Generally, RREA simulations can provide ever larger enhancements than observed uniquely on the mountain tops because the lightning initiation process with the following potential difference brake is not included in the codes and remains unknown. Naturally, TGEs reaching significant flux are abruptly terminated by lightning flashes. Rarely certain atmospheric conditions may prevent lightning initiation at high altitudes near mountain tops, allowing the RREA process to develop longer than usual.

Figures 5-7 provide detailed information on TGE development measured by the STAND1 unit near the GAMMA surface array. We demonstrate count rates of 1-cm thick upper and stand-alone 3-cm thick scintillators. The energy threshold of 1 cm thick scintillator is lower than 3 cm thick. Thus, its count rate is higher. The STAND1 unit near the SKL hall was buried in heavy, wet snow, causing TGE particle absorption. As we see in the Figures, the count rate was rather stable at the minute of maximum flux on one 1-minute time scale (Fig. 6a) and 50 ms time scale (Fig. 7a). Then we calculate the count rate mean values and variances before TGE at fair weather at the same time one day before (Figs 6b and 7b). As we see from the figures, the count rate enhanced more than ten times. The TGE significance recalculated for a minute time series (a standard adopted for the TGE comparisons) for the 1 cm thick scintillator reaches ≈1000% and ≈1500 standard deviations from the background mean value. Comparing the relative errors of the TGE count rate in 1 cm thick and the same at fair weather, it was found that the count rate errors were 3% and 4.4%, respectively. The maximum minute of TGE showed a more stable particle flux than the ambivalent cosmic ray flux observed during fair weather. The same behavior is found for the 3 cm thick scintillator and the 50 ms time series.

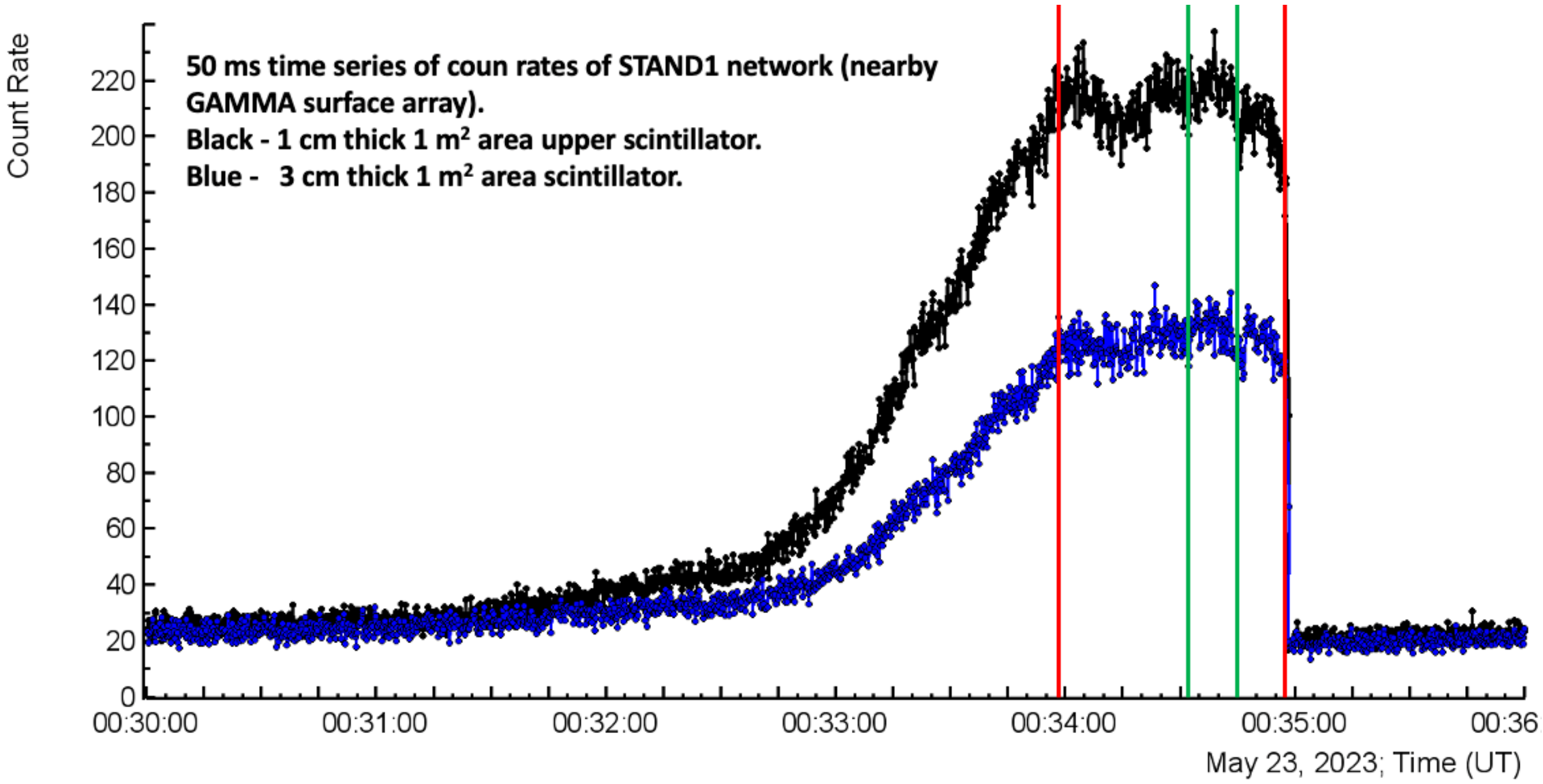

**Figure 5.Development of 23 May TGE. By the red lines, we show the maximum flux minute of TGE, and by the green lines – the maximum flux is 10 seconds.**

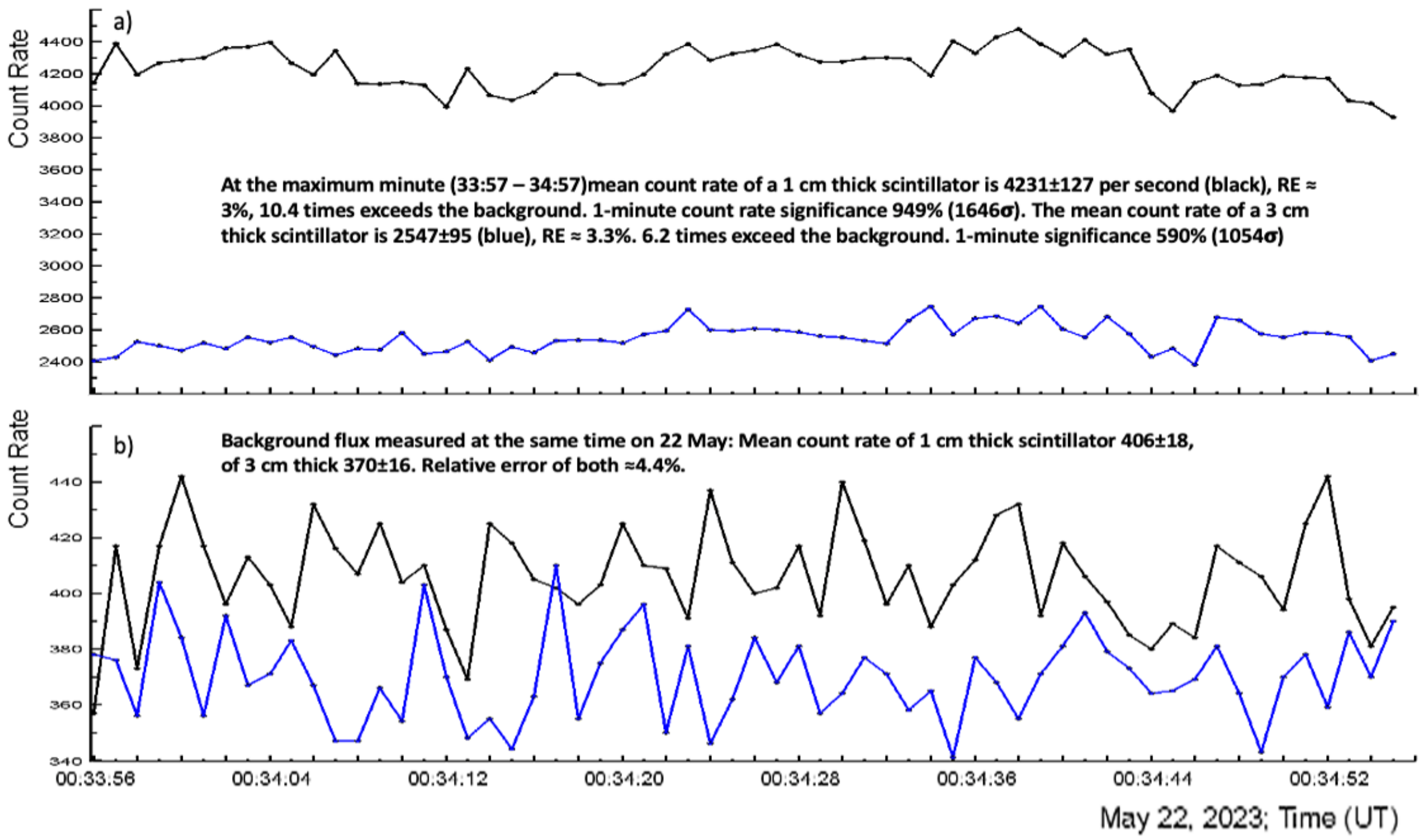

**Figure 6. a) - the maximum minute of the TGE, b) – the background count rate.**

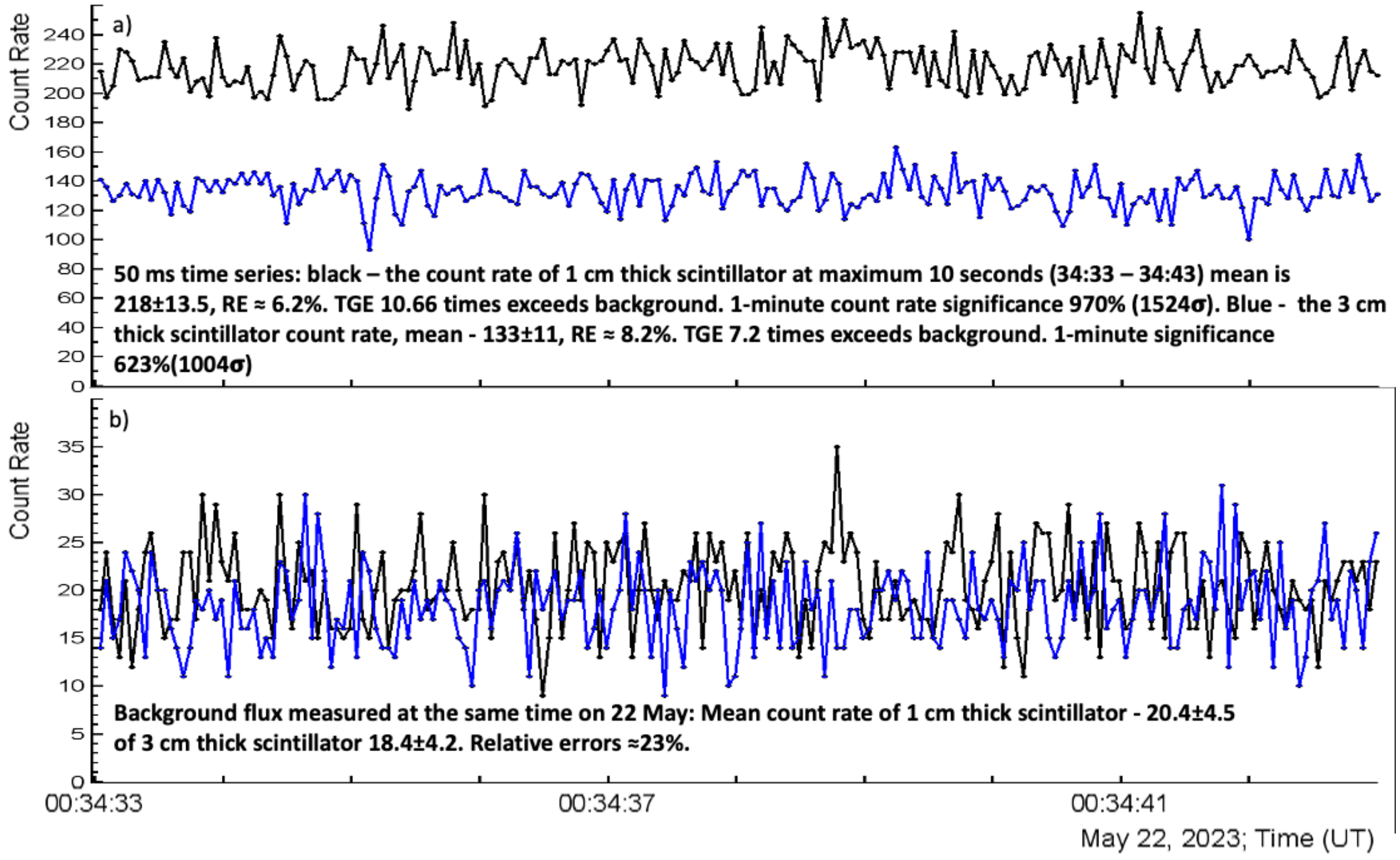

**Figure 7. a) time series of the maximum 10 seconds of TGE, b) background count rates**

## 4. Digitization of particle signals from the fast oscilloscope

The digitizing oscilloscope sampling rate of 6.4 ns is smaller than the particle pulse full width on a half maximum (FWHM) of 25-35 ns. To find genuine particles from the oscilloscope raw data and avoid double counting, we are looking for the genuine particle pulse and suppress false counting due to double pulses. For it, we calculate the derivatives computed from the sequence of sampling amplitudes. When the derivative sign changes from negative to positive and the amplitude is lower than 32 mV, we consider the sampling time as a particle arrival, see Fig.8. If we look only at amplitudes, we will count five particles below the threshold of -32 mV; however, there is only one! Thus, the time stamp accuracy is 6.8 ns for the 1-cm thick CUBE scintillators in the SKL experimental hall and four ns for the 3-cm thick scintillator in the MAKET experimental hall.

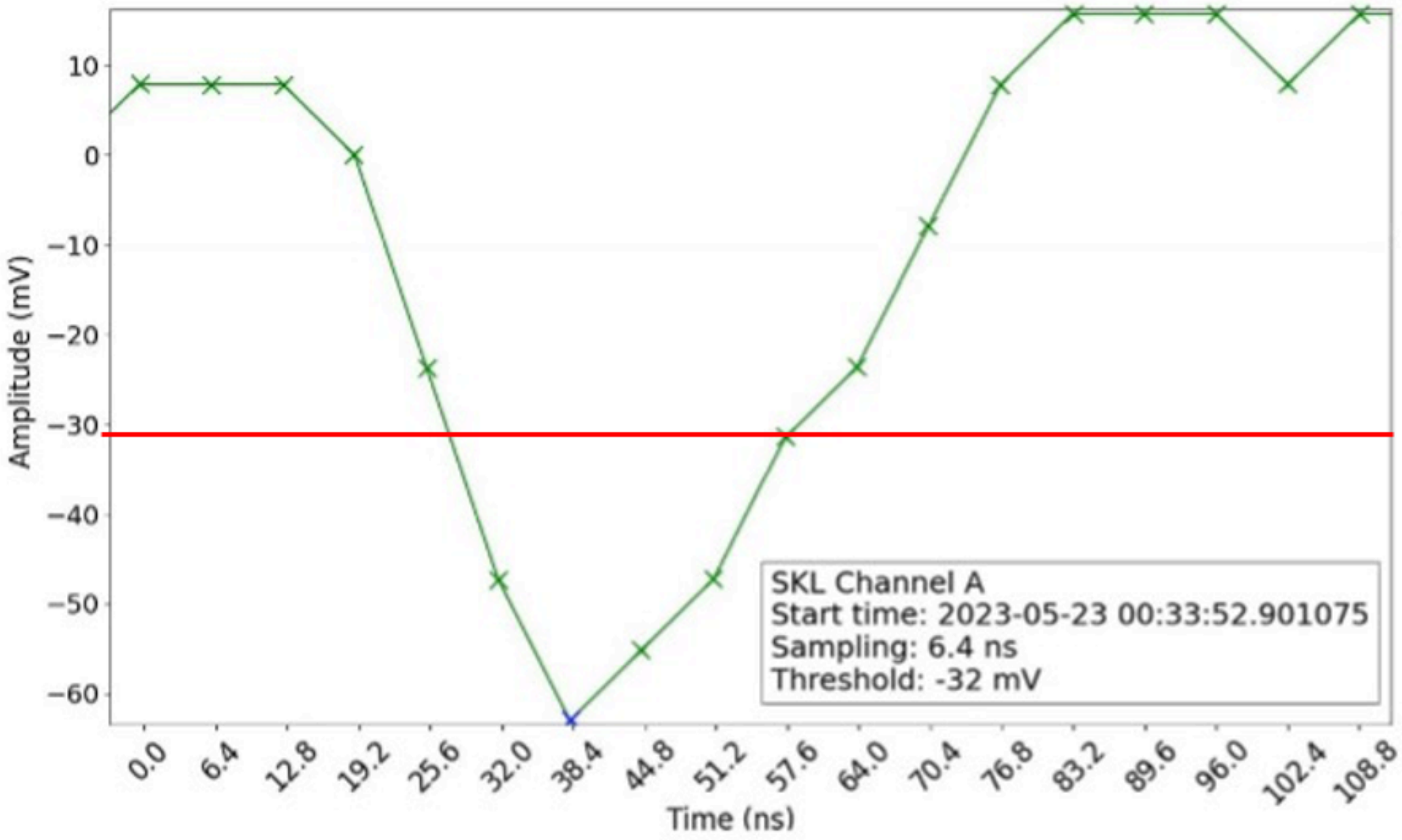

**Figure 8.The shape of the pulse as the oscilloscope enumerates it. The pulse is distributed among several bins of 6.4 ns each, shown as crosses. The red line shows the pulse amplitudes below the threshold of 32 mV. The full width on a half maximum of the pulse is ≈30 ns.**

Fig. 9 presents distributions of pulse amplitudes measured during TGE and at fair weather, 100 ms before trigger at 00:33:52.9 and 100 ms after trigger (the same sample for the fair weather). In the 3,125 million sampling intervals, 331 particles were selected during TGE and 123 particles at fair weather (CUBE's N2 scintillator attached to oscilloscope's channel A). The difference is 208 for 200 ms, i.e., expected 1040 for one second.

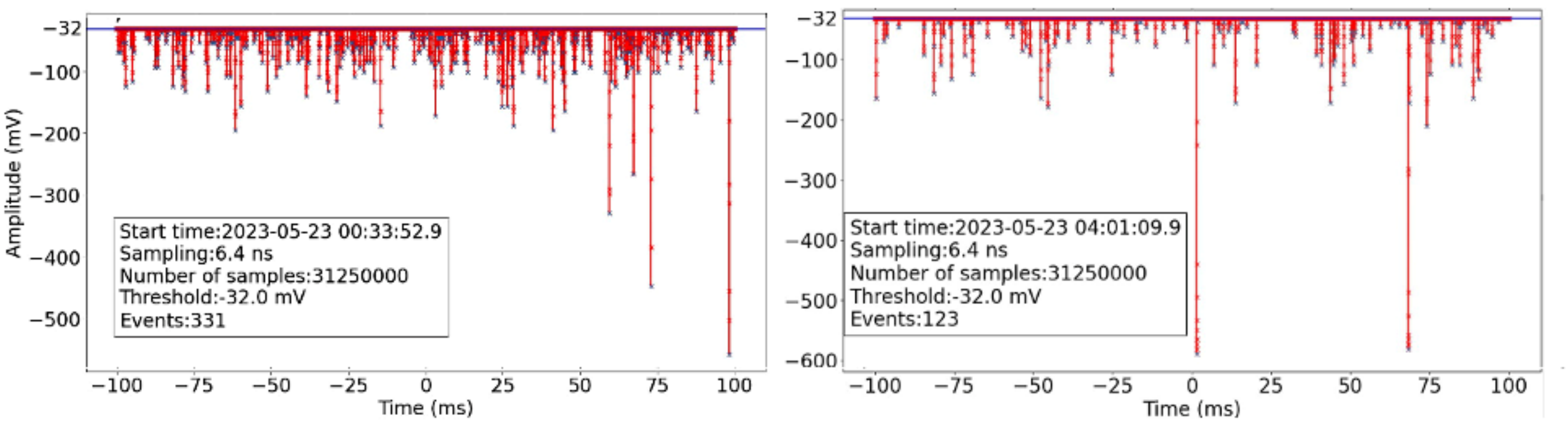

**Figure 9. Signals registered by the digital oscilloscope (channel A, CUBE 2). The oscillogram contains data for 200 ms, 100 ms before trigger, and 100 ms after trigger that occurred at 00:33:53 UT on May 23, 2023, frame a), and at 04:10 the same day, frame b). In the insets, we show signal sampling information and the number of selected events.**

The tradeoff between suppressing the PM noise and obtaining high signal detection efficiency leads to setting the threshold of the input signal stream to -32 mV. However, we have to check if the count rate is not suppressed by choosing this threshold. As we can see from Figure 2, the three CUBE scintillators (N 2,3,6) are attached directly to the oscilloscope. To control their count rate, we use the same type of detector, CUBE's N 1, attached to the pulse analyzer. Both detectors are located nearby on the floor of the SKL hall, see Fig.3. We check the count rates of all detectors by attaching one by one to the analyzer and comparing count rates. Figure 11 shows the 1-minute count rate of the CUBE's scintillator N1 during TGE. At the maximum minute of TGE, the count rate was 102,500, the difference with fair weather count – 69,000, which makes 1150 per second. It is rather close to the number expected for the 1-second difference for the CUBEs' second detector – 1040. In the inset to Fig.8, we can see that CUBE's 2,3,4 scintillators' 200 ms count rate coincide rather well. Slightly different efficiencies of scintillators can explain the small in-channel differences.

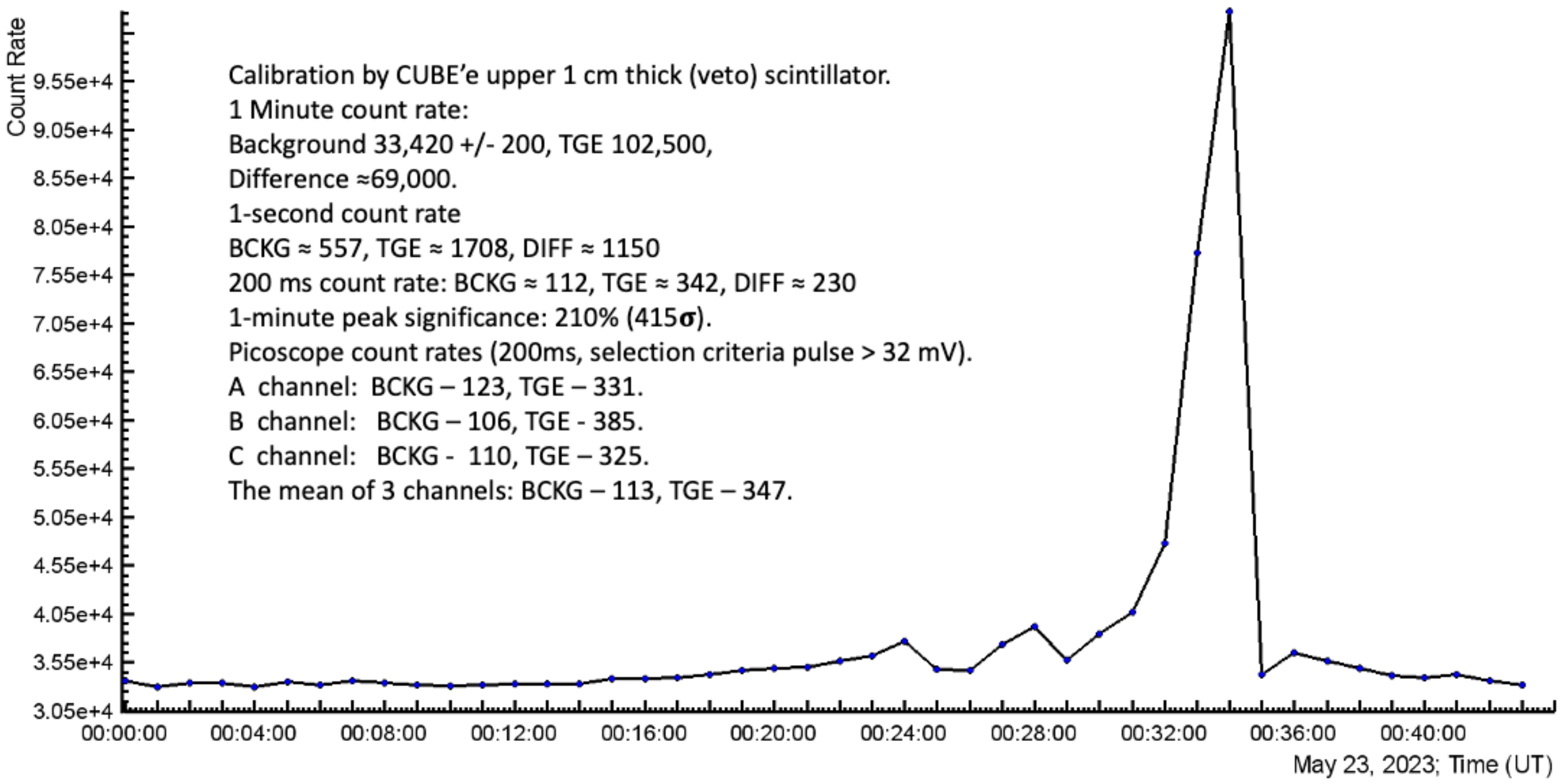

**Figure 10. Time series of 1-minute count rate of 1 cm thick plastic scintillator CUBE 1. The inset compares recalculated 200 ms count rates of the first (veto) CUBE's scintillator with A, B, and C oscilloscope channels (CUBE 2,3,6).**

The pulses from the outdoor 3-cm thick scintillator near the MAKET experimental hall are larger; see Fig.11, a, b, c, and d. Also, we detect double-pulsed patterns (see Fig. 12). To avoid double counting, we set a "dead-time" of 1s, merging neighboring pulses and counting them as one.

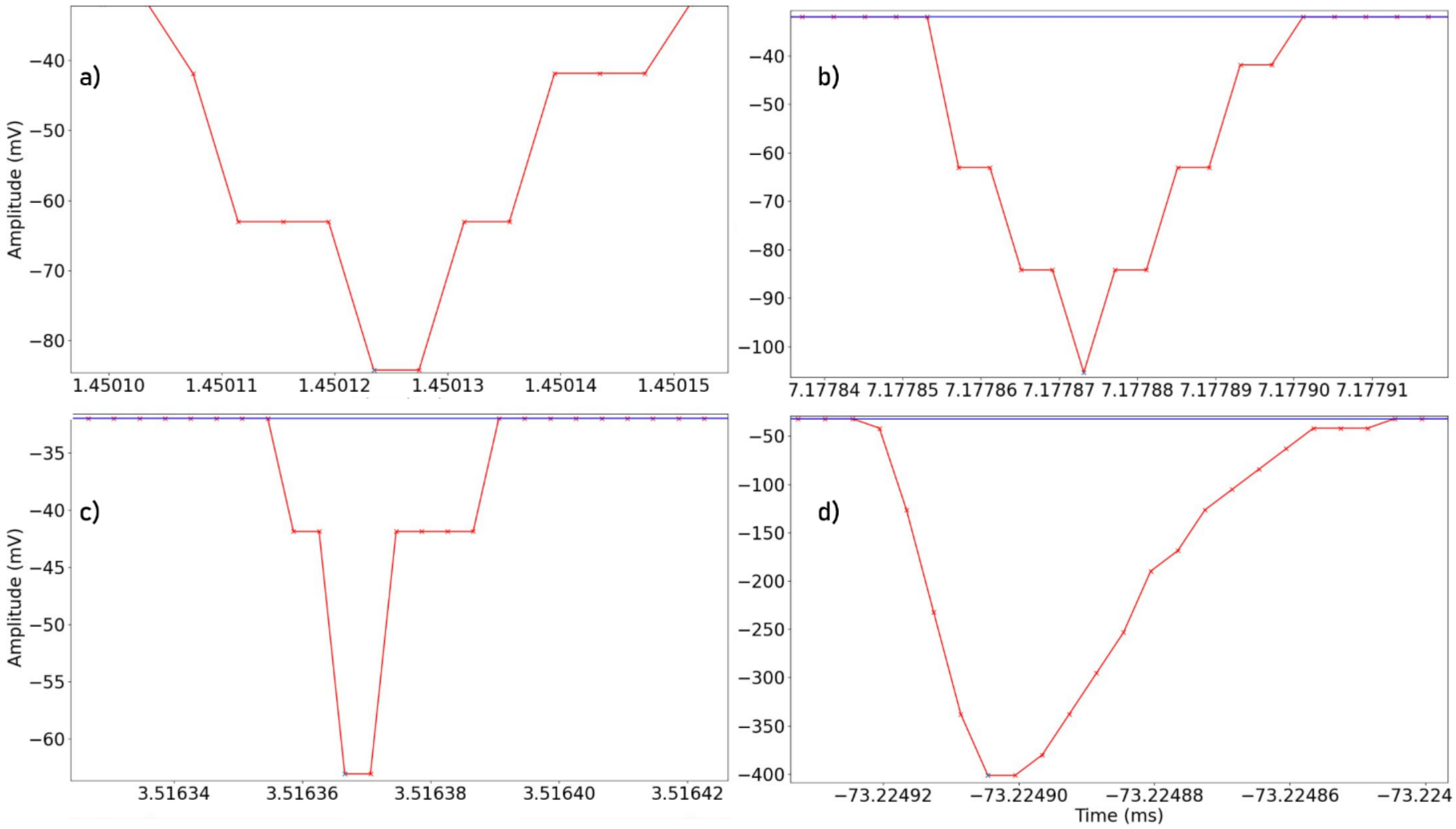

**Figure 11. Different shapes of the signals from the 3 cm thick scintillator digitized by the oscilloscope, sampling time four ns.**

**Figure 12. Neighboring pulses will be counted as two particles; we introduce a “dead time” of 1 □s to avoid double counting.**

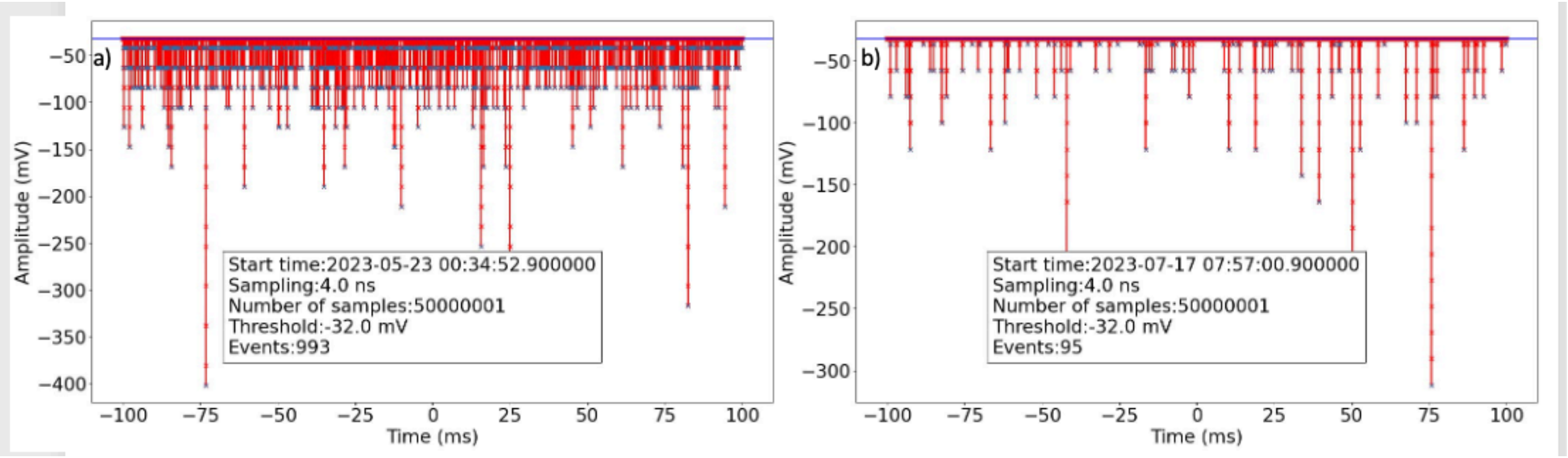

**Figure 13. Signals registered by the digital oscilloscope (channel B, 3 cm thick stand-alone outdoor scintillator). The oscillogram contains data for 200 ms, 100 ms before trigger, and 100 ms after trigger that occurred at 00:34:52.9 UT on May 23, 2023, frame a) and at 07:57:01 on July 17, 2023, frame b). In the insets, we show signal sampling information and the number of selected events.**

## 5. Analysis of time distribution between successive signals for the SKL and MAKET time series and correlation analysis of the three channels of SKL signals on 23 May 2023 TGE.

The distribution of particle arrival times can inform on the electron-photon avalanche development. The ECSs initiated by a single seed electron possibly change the temporal pattern of the particle arrival, making it different from the expected Poisson distribution. We look for any significant differences from the Poisson distribution of the particles registered within a time interval ΔT and compare it with the same number of events generated from the Poisson distribution. We use the mean value from 10 independent samples from the Poisson distribution. The same procedure was applied to the data obtained on fair weather. In the second column, we put the intensity of particle flux recalculated from the count rate in 200 ms; in the third, the time window in which we are looking for the coincidences; in the fourth - number of events falling in this window; and in the fifth – the same number but obtained with modeled Poison variables. The symmetric right side of the Table shows the same parameters but obtained in fair weather.

**Table 1. Comparison of observed time between successive events (fallen within a chosen period) with expected from the Poisson distribution for the TGE and fair weather.**

| TGE: SKL (A, B, C channels,) and MAKET (2 triggers, B channel) | | | | | Fairweather (same channels as for TGE) | | | |
|---|---|---|---|---|---|---|---|---|
| | $I_0$ | ΔT (□s) | Events in ΔT | Poisson events in ΔT | $I_0$ | ΔT (□s) | Events in ΔT | Poisson events in ΔT |
| Channel A | 1655 | 0-10 | 7 | 5±2.2 | 615 | 0-10 | 1 | 0.75±0.87 |
| | | 10-100 | 42 | 45±6.7 | | 10-100 | 6 | 6.5±2.6 |
| | | 100-500 | 146 | 136±11.7 | | 100-500 | 23 | 25±5 |
| Channel B | 1925 | 0-10 | 15 | 7±2.6 | 530 | 0-10 | 0 | 0.56±0.76 |
| | | 10-100 | 36 | 60±7.7 | | 10-100 | 4 | 4.8±2.2 |
| | | 100-500 | 181 | 170±13 | | 100-500 | 23 | 19±4.3 |
| Channel C | 1625 | 0-10 | 11 | 5±2.2 | 550 | 0-10 | 6 | 0.66±0.78 |
| | | 10-100 | 34 | 43±6.6 | | 10-100 | 9 | 5.3±2.3 |
| | | 100-500 | 134 | 131±11.4 | | 100-500 | 25 | 21±5 |
| MAKET | 3020 | 1-10 | 15 | 18±4.2 | 475 | 0-10 | 2 | 0.44±0.66 |
| | | 10-100 | 151 | 139±11.8 | | 10-100 | 3 | 4±2 |
| | | 100-500 | 292 | 312±17.7 | | 100-500 | 16 | 15.6±4 |
| MAKET | 4960 | 1-10 | 35 | 48±6.9 | 475 | 0-10 | 2 | 0.44±0.66 |
| | | 10-100 | 342 | 340±18.4 | | 10-100 | 3 | 4±2 |
| | | 100-500 | 540 | 521±22.8 | | 100-500 | 16 | 15.6±4 |

As indicated in the table, there are differences between the expected Poisson values and the actual measurements in small time frames, particularly in channels B and C. These deviations could be attributed to ECS particles with minimal time separation. However, it is also possible that the tails of extensive air showers cause the arrival of time-tight events. Furthermore, we observe a significant discrepancy in channel C during fair weather conditions.  No significant differences exist in the experimentally measured distances between successive events and Poison generated samples for the large time windows from 100 to 500 □s.

Table 2 compares the inter-channel correlations with the expected Poisson values. To do this, we generated ten independent samples following the Poisson distribution. We compared the mean number of events within time windows ranging from 320 ns to 640 □s to the measured values. We found discrepancies in the correlations between A and B channels at smaller time windows. However, the other two channels did not show any significant differences.

**Table 2. The correlation analysis of registered signals from SKL plastic scintillators**

| Coincidences SKL | $I_0$ | ΔT (□s) | Number of coincidences (Poison) in ΔT | Number of coincidences SKL A B C |
|---|---|---|---|---|
| AB | 1655 | 0.32 | 1.0 | 5 |
| | 1925 | 0.64 | 2.0 | 15 |
| | | 1.28 | 4.4 | 20 |
| | | 6.4 | 20.1 | 20 |
| | | 12.8 | 39.9 | 45 |
| | | 128 | 402.7 | 390 |
| | | 640 | 2027.0 | 2055 |
| AC | 1655 | 0.32 | 0.9 | 0 |
| | 1625 | 0.64 | 1.8 | 5 |
| | | 1.28 | 3.5 | 5 |
| | | 6.4 | 17.3 | 15 |
| | | 12.8 | 34.7 | 30 |
| | | 128 | 339.9 | 325 |
| | | 640 | 1719.6 | 1760 |
| BC | 1925 | 0.32 | 1.1 | 0 |
| | 1625 | 0.64 | 1.9 | 0 |
| | | 1.28 | 3.9 | 0 |
| | | 6.4 | 20.6 | 15 |
| | | 12.8 | 39.1 | 50 |
| | | 128 | 400.5 | 365 |
| | | 640 | 2004.3 | 2020 |

## 6. Uniformity of TGE signals, comparison with TGF data

Recent Atmosphere-Space Interactions Monitor (ASIM) measurements on board the International Space Station provide a vast amount of statistically provided TGFs. ASIM detector is much more efficient in registering gamma-ray bursts than other orbiting gamma laboratories designed to detect gamma rays from violent explosions in the Universe and use complicated off-line triggers to find TGFs from the Earth's direction. This allows for directly comparing the millisecond time-scale time series of TGEs and TGFs. In Fig 14, we compare the 10 ms duration time series of the multi-pulse TGF [25], registered by ASIM, and the same duration time series from the STAND1 detector. The difference between TGFs, which last for milliseconds, and TGEs, which last for minutes, is due to the distance between the particle source and the detector. TGEs are detected ≈1000 times closer to the source than TGFs. Only the most energetic gamma rays from RREAs can reach the Space Station 400 km from Earth, producing a short particle burst. Sometimes, multi-pulse TGFs are registered by ASIM (refer to figures in [25]). On Aragats, RREAs produce a nearly continuous flow of particles for minutes due to the proximity to source detectors exposed to the particle flux. The light pulse from ASIM's BGO is several hundreds of ns long; therefore, it has an effective time resolution of about 1 μs and a size of 900 $cm^2$. The STAND1 detector (size is 10,000 $cm^2$ ) with a time resolution of 20-30 ns is attached to a digitizing oscilloscope with a sampling rate of 4 ns. Both detectors register 50 particles in 10 ms. However, the particle flux density at ASIM is ≈ ten times larger due to its smaller size. Fig 14a shows a very intense gamma-ray burst (left side) followed by a few discrepant gamma rays. In Fig 14b distribution of particles is more-or-less uniform.

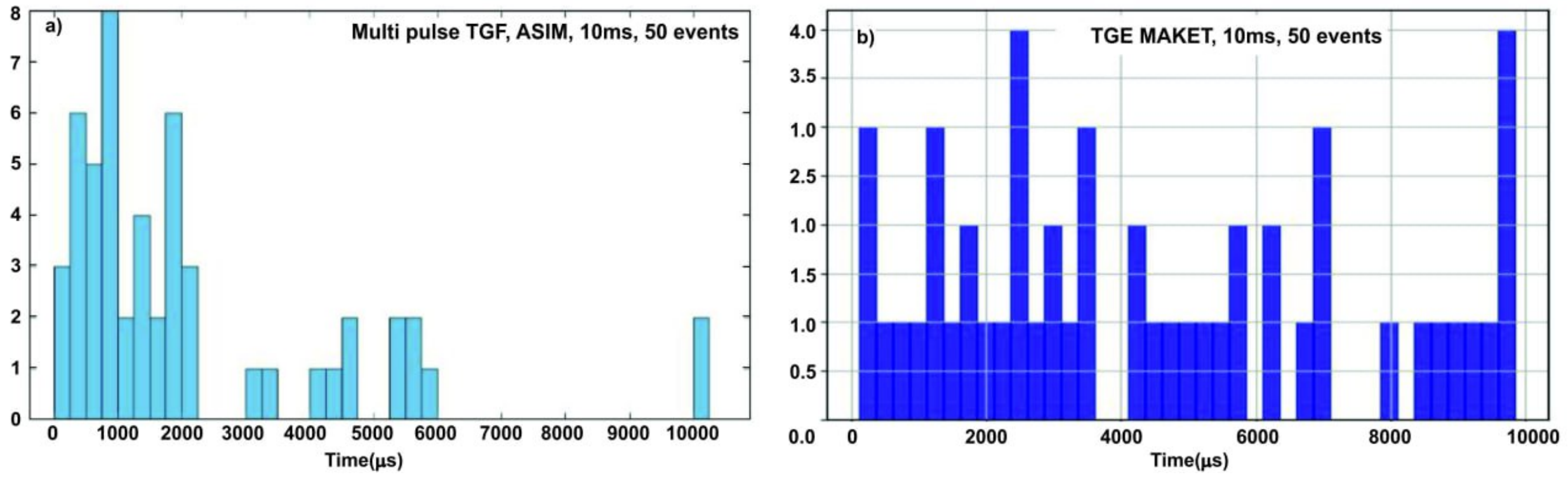

**Figure 14. a) Time series of gamma-ray arrival times from the onset of the first TGF until the onset of the last TGF, registered by ASIM's HED detector; b) Time series of the TGE particle arrival times registered by STAND1 (MAKET); zero time corresponds to 00:34:53.1 on May 23, 2023. The bin width of both is 250 □s.**

Figure 15a shows the extended to 200 ms time series of the STAND1's TGE detection. Figure 15b shows a time series obtained with randomly generated 992 events (the same number of events as in 16a) from the Poisson distribution. The simulation aims to demonstrate that TGE particles come randomly from multiple avalanches in a large-scale thunderous atmosphere. Instead, TGFs comprise a few most energetic avalanches, occasionally reaching the fast-moving detector at distances 400 km from the source.

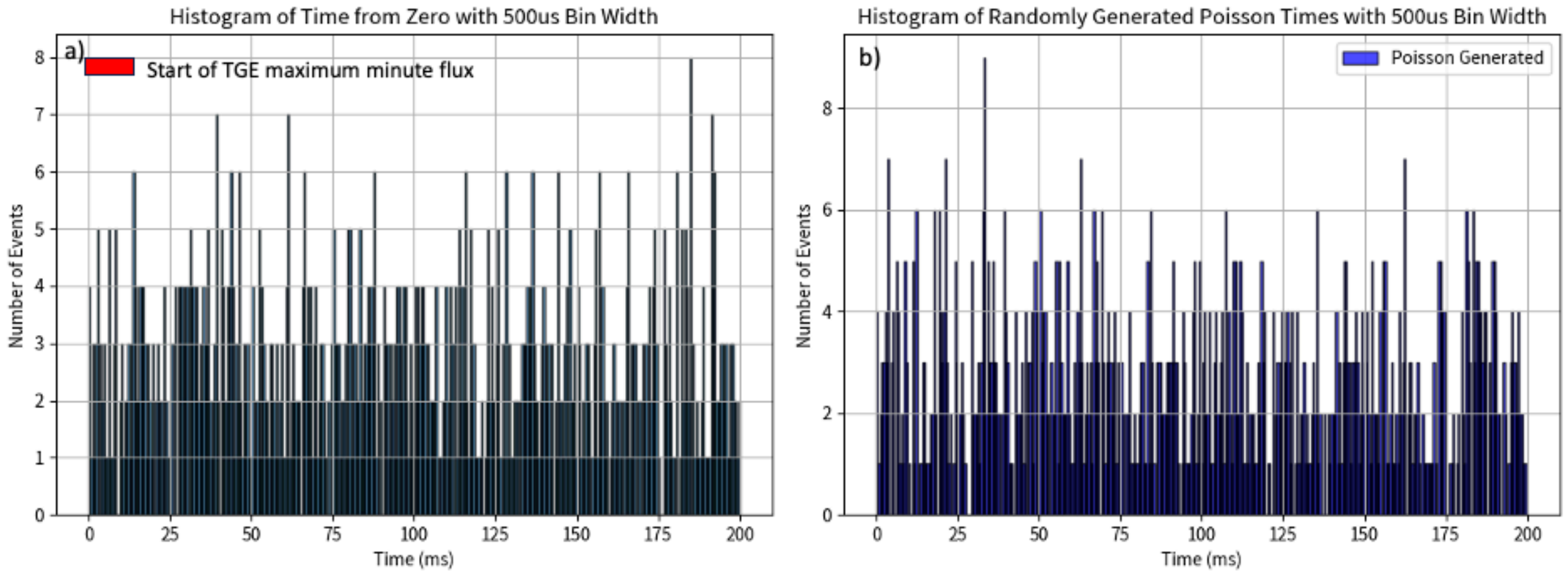

**Figure 15 a) 200 ms time series of the TGE particle arrival times registered by STAND1 (MAKET), zero time corresponds to 00:34:52.9 on May 23, 2023. b) time series of the Poisson random times within 200 ms. The bin width of both is 500 □s.**

In Figure 16, we demonstrate the uniformity of particle arrival over 200 ms of recorded time series. The frames (a-d) show successive 10 ms long time series of 50 registered particles with ordered numbers from 700 to 900. These time series are similar and do not show enormous bursts that could be attributed to very energetic ECS occasionally captured by the detector, as in the ASIM case.

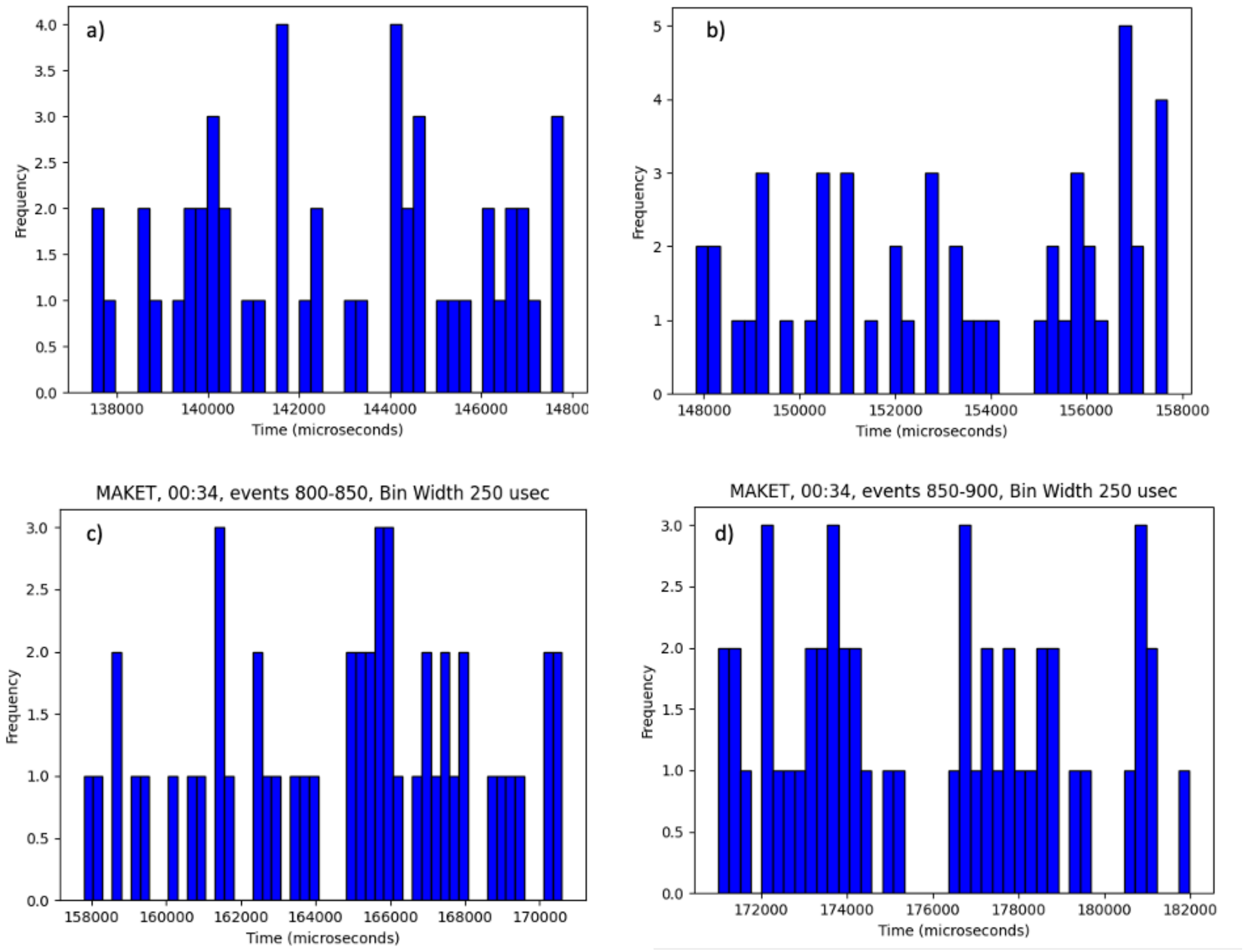

**Figure 16. Successive 10 ms time series of TGE particle detection by MAKET 3-cm thick scintillator attached to digitizing oscilloscope.**

To prove that the arrival time of TGE particles follows exponential interarrival time distribution (directly connected to Poisson distribution), we conducted a Kolmogorov-Smirnov (KS) test. Our approach involved generating 100 independent samples, each containing 992 timestamps, distributed according to Poisson distribution within a 200ms interval. We then sorted these timestamps in ascending order and calculated the time difference between successive numbers. Figure 17 displays the pairwise comparisons of the MAKET data at 00:34:52.9 with one of 100 generated Poisson-distributed samples.

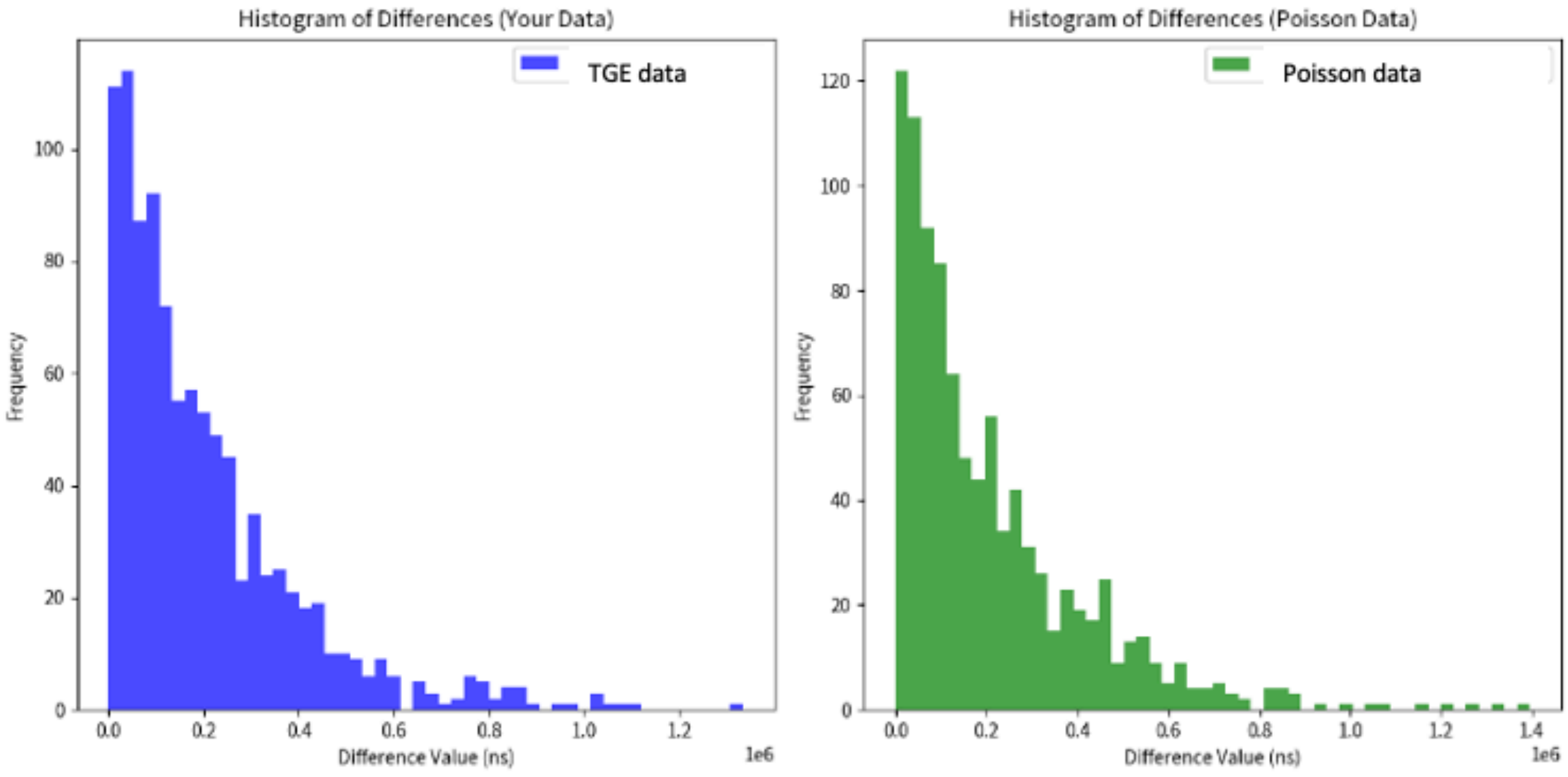

**Figure 17. Histogram of the KS test statistics for pair-wise comparison of MAKET and 100 Poisson distributed samples.**

The mean value of the KS test averaged by 100 values was 0.036, corresponding P-value of 0.544. Based on the high P-value of the KS statistics, it is evident that TGE particle arrivals occur independently of one another on May 23, 2023, and the time intervals between their arrivals follow a Poissonian process.

## 7. CORSIKA simulation of TGE particle arrival times and distances

To get insight into the expected space-temporal distributions of the RREA particles, we perform a simulation study with the CORSIKA code [26] version 7.7400, which considers the electric field's effect on the particles' transport. We use electrons with energy 1 MeV as seed particles; the electric field was introduced at 5200 m in a strong electric field of 2.1 kV/m. We store arrival times and radial distances of all RREA electrons reaching the ground level of 3200m from 1000 simulation trials. The propagation of electrons and gamma rays was followed in the avalanche until their energy decreased to 0.05 MeV. The actual energy threshold depends on the particular particle detector used in measurements. In our simulations, we have set a 4 MeV energy threshold for the indoor plastic scintillators. As illustrated in Figure 18a, the mean arrival time of RREA electrons is approximately 7.5 µs. In contrast, the rough estimate of the electron collection time, Full Width at Half Maximum (FLHM), is around 200 ns. For radial distances shorter than 10 m, the RREA collection time is even smaller - approximately 150 ns (Fig. 18b).

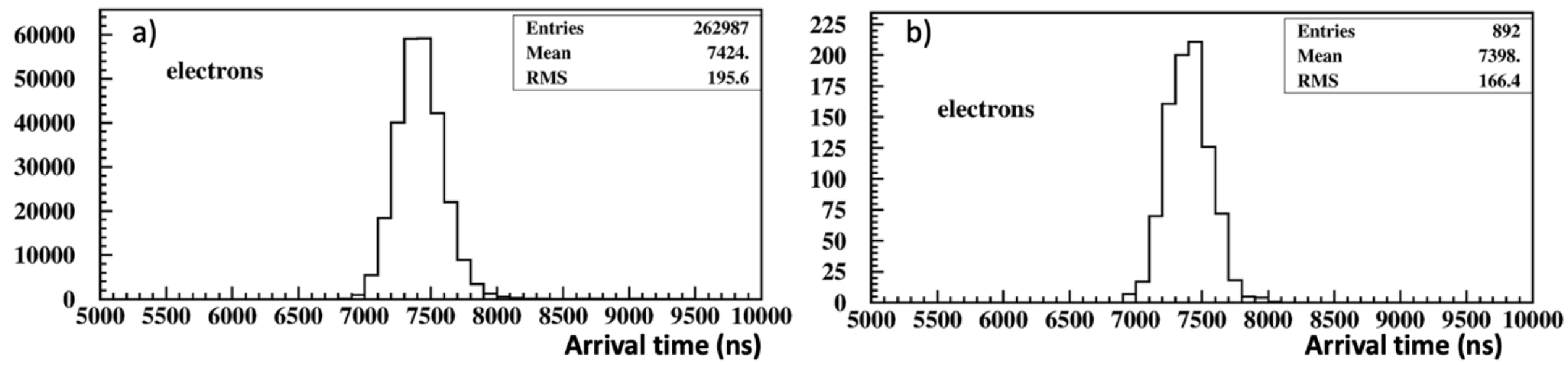

**Figure 18. The *arrival time* distribution of *electrons* total a) and a subsample in 10 m radii *b*).**

A large number of secondary electrons are produced in the atmospheric electric field and arrive at the Earth's surface in a very short time. The strength of the electric field is approximately 25% higher than the critical strength. Although the seed electrons are injected into the electric field from the same point, the radial distribution is quite dispersed. Only about 3% of the electrons fall within a circle with a radius of 10 meters around the injection point projected onto the ground.

**Discussion and conclusions**

We have conducted a space-temporal analysis of the RREA process within the thunderous atmosphere. This analysis is based on detecting large TGE on May 23, 2023, by a STAND1 distributed network of particle detectors. Our survey of the TGE includes a comprehensive statistical analysis of time series ranging from nanoseconds to seconds (Figs 5,6,7, 15). The arrival of RREA particles, consisting of electrons and gamma rays with energies greater than 4 MeV, at the Earth's surface is uniform. Statistical analyses (Tables 1 and 2) of particle arrival times and spatial correlations do not provide conclusive evidence for detecting separate ECSs (microbursts). Rather, TGE comprises a mixture of thousands of ECSs born in the large-scale electric field. The information regarding their origin from a single ECS, which originates from a seed electron on a particular height, is completely smeared due to the large spatial dispersion of millions of secondary particles. The simulation study confirms our conclusion. The distribution of particles over several square kilometers results in rather low particle densities, even at the maximum TGE flux density is approximately 0.5 per square centimeter per second. The statistical analysis of the space-temporal pattern of particle arrival indicates a diffuse, not intense, "rain-type" flux of electrons and gamma rays covering large areas on the Earth's surface.

To sum up, the RREA in the thundercloud above Aragats produced a stable particle flux on 23 May 2023 that was ten times higher than the fair weather ambient cosmic ray flux. The electron accelerator in the thunderclouds generates particle fluxes, which are detected as TGEs on the ground. The count rate of TGEs shows a smaller relative error when compared to the flux of the ambient population of cosmic rays during fair weather. TGE particles arrive randomly, according to the Poisonian process.

**ACKNOWLEDGMENTS**

The authors would like to express their gratitude to the Aragats Space Environmental Center staff for ensuring the uninterrupted operation of all particle detectors and field meters. We would also like to thank Y. Khanikyanc for maintaining the file recording infrastructure and S. Soghomonyan's valuable contributions to discussions on the initial project stage. Additionally, A.C. would like to thank Davit Aslanyan for preparing Figure 4. The authors acknowledge the support of the Science

Committee of the Republic of Armenia (Research Project No. 21AG-1C012) in the modernization of the technical infrastructure of high-altitude stations.